\documentclass{iaus}
\usepackage{graphicx}

\title[Magnetic field and circumstellar environment of HD 57682] 
{A MiMeS analysis of the magnetic field and circumstellar environment of the weak-wind O9 sub-giant star HD 57682}

\author[J.H. Grunhut et al.]   
{J.H. Grunhut$^{1}$, G.A. Wade$^1$, W.L.F. Marcolino$^2$, V. Petit$^3$, and the MiMeS Collaboration}

\affiliation{$^1$Kingston, Canada; $^2$Marseille, France; $^3$West Chester, USA\\}

\pubyear{2010}
\volume{272}  
\pagerange{119--126}
\jname{Active OB stars}
\editors{A.C. Editor, B.D. Editor \& C.E. Editor, eds.}
\begin{document}

\maketitle

\begin{abstract}
I will review our recent analysis of the magnetic properties of the O9IV star HD\,57682, using spectropolarimetric observations obtained with ESPaDOnS at the Canada-France-Hawaii telescope within the context of the Magnetism in Massive Stars (MiMeS) Large Program. I discuss our most recent determination of the rotational period from longitudinal magnetic field measurements and H$\alpha$ variability - the latter obtained from over a decade's worth of professional and amateur spectroscopic observations. Lastly, I will report on our investigation of the magnetic field geometry and the effects of the field on the circumstellar environment.
\end{abstract}

\firstsection 
\section{Introduction}
The presence of strong, globally-organized magnetic fields in hot, massive stars is rare. To date, only a handful of massive O-type stars are known to host magnetic fields. In 2009, Grunhut et al. reported the discovery of a strong magnetic field in the weak-wind O9IV star HD 57682 from the presence of Zeeman signatures in mean Least-Squares Deconvolved (LSD) Stokes $V$ profiles. Their analysis of the $IUE$ and optical spectra determined the following atmospheric and wind properties: $T_{eff}=34.5$\,kK, $\log(g)=4.0\pm0.2$, $R=7.0^{+2.4}_{-1.8}$\,$R_{\odot}$, $M=17^{+19}_{-9}$\,$M_{\odot}$, and $\log({\dot{M}})=-8.85\pm0.5$\,$M_{\odot}$\,yr$^{-1}$.



\section{Temporal Variability}
\begin{figure}
\centering
\includegraphics[width=5.3in]{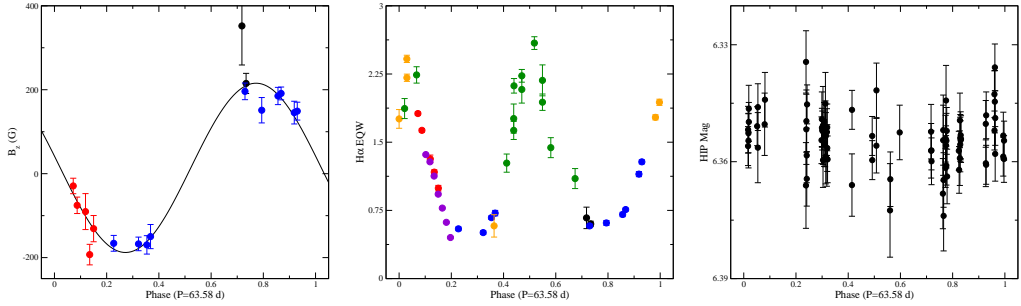}
\caption{Phased longitudinal magnetic field measurements (left), H$\alpha$ equivalent width variation (middle), and Hipparcos photometry (right), for HD 57682. Different colours indicate different epochs of observations.}
\label{sidebyside}
\end{figure}

Both the longitudinal magnetic field and H$\alpha$ equivalent width of HD 57682 are strongly variable. In addtion to our 17 ESPaDOnS observations, we've also utilized H$\alpha$ observations from amateur spectroscopy from the BeSS database, as well archival ESO UVES and FEROS observations dating back over a decade. A period search of these data resulted in a period of $\sim$31\,d, consistent with the rotational period estimated by Grunhut et al. (2009). However, the magnetic data could not be reasonable phased with this period. Ultimately, adopting a period of 63.58\,d (twice the period obtained from the H$\alpha$ data) resulted in a coherent phasing of all the data at our disposal, as shown in Fig.~\ref{sidebyside}.

The longitudinal magnetic field appears to vary sinusoidally, consistent with a magnetic field dominated by a strong dipolar component. The H$\alpha$ equivalent width shows a double-wave pattern with peak emission occurring at the magnetic crossover phases (i.e. when the longitudinal field is null).

The photometric light curve from Hipparcos shows no apparent variability. This likely indicates that the column density of the magnetically confined plasma is relatively low at eclipse phases.

\section{Magnetic Geometry and Circumstellar Environment}
\begin{figure}
\centering
\includegraphics[width=2.2in]{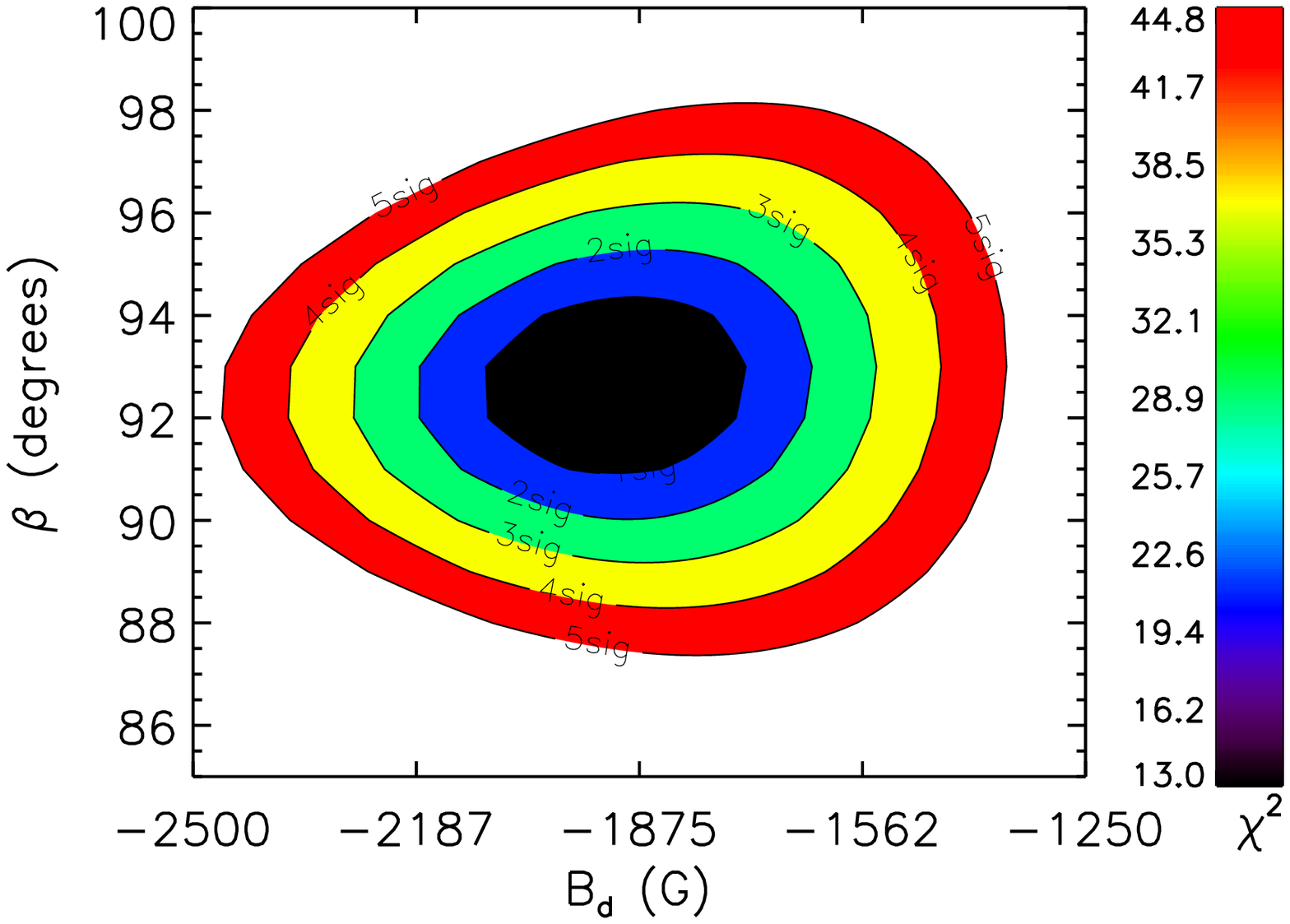}
\includegraphics[width=1.9in]{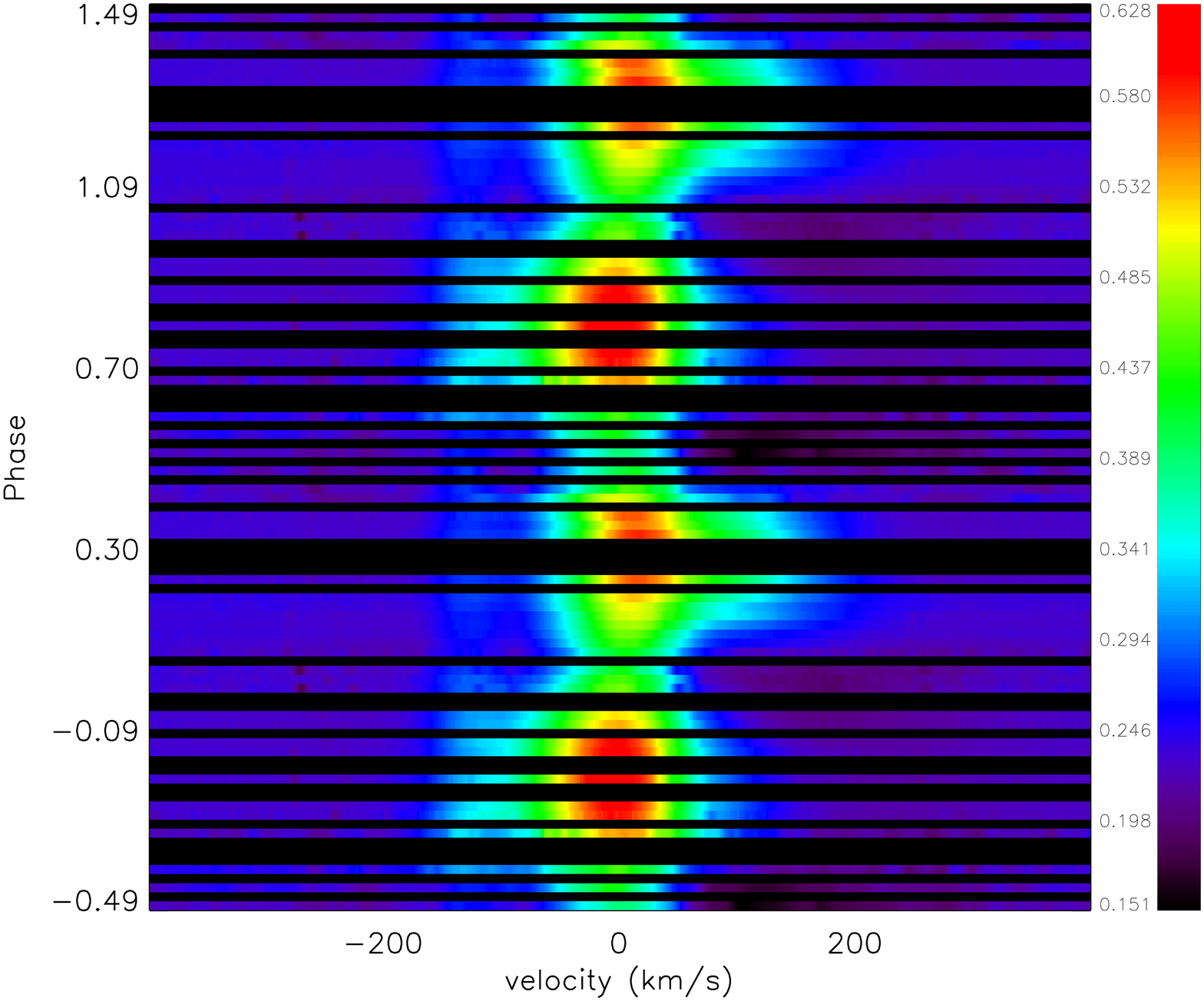}
\caption{{\bf Left:} $\chi^2$ landscape as a function of dipole field strength ($B_d$) and obliquity angle ($\beta$). Shown are the intervals corresponding to 1, 2, 3, 4, and 5$\sigma$ using $i=22^{\circ}$. {\bf Right:} Phased H$\alpha$ residual variations relative to an LTE model. Note the variability is likely due to a magnetically confined wind.}
\label{chisq}
\end{figure}

Using the longitudinal component of the magnetic field, measured from the mean LSD Stokes $V$ and $I$ profiles, we are able to fit a dipole model, characterized by the magnetic field strength at the poles ($B_d$) and the angle of obliquity of the magnetic axis relative to the rotation axis ($\beta$). Using the $v\sin i$ and radius as determined by Grunhut et al. (2009) and assuming rigid rotation, we can infer an inclination angle $i=22^{+17\,\circ}_{-9}$. Using this value we obtain the $\chi^2$ landscape as a function of $B_d$ and $\beta$ shown in Fig.~\ref{chisq}. Taking into account the range of possible inclination angles, we find that $B_d=1-3$\,kG and $\beta=94^{\circ}\pm5^{\circ}$.

In Fig.~\ref{chisq} we also show the residual variations of H$\alpha$ phased with the adopted rotational period. We conclude based on the characteristics of this dynamic spectrum that the magnetic field is exerting strong confinement (confinement parameter $\eta*\sim10^3-10^5$; ud-Doula \& Owocki 2002) on the weak wind of HD 57682, resulting in the H$\alpha$ variability. However, the slow rotation is likely unable to centrifugally support a stable magnetosphere. Therefore, the plasma that is present likely has a relatively short residence time in the magnetosphere, and must therefore be continually replenished (see Townsend et al. these proceedings).

\end{document}